\documentstyle[a4,12pt,epsf]{article}
\newcommand{\noi}{\noindent}
\newcommand{\eq}{\begin{equation}}
\newcommand{\en}{\end{equation}}
\newcommand{\eqa}{\begin{eqnarray}}
\newcommand{\ena}{\end{eqnarray}}

\newcommand{\eff}{e \! f \! f}

\newcommand{\tzeta}{{\tilde \zeta}}

\newcommand{\cak}{ {\cal K}}

\newcommand{\cao}{ {\cal O}}

\newcommand{\bpartial}{{\bar \partial}}

\newcommand{\tG}{{\tilde G}}

\newcommand{\ageq}{\mbox{}_{\textstyle \sim}^{\textstyle > }}

\newcommand{\Ra}{\Rightarrow}

\begin{document}

\renewcommand{\baselinestretch}{1.1}
\renewcommand{\thefootnote}{\arabic{footnote}}
\setcounter{footnote}{0}
\renewcommand{\theequation}{\arabic{section}.\arabic{equation}}
\renewcommand{\thesection}{\arabic{section}}
\language0

\hbox{}
\noindent April 2000 \hfill

\vspace{0.5cm}
\begin{center}

{\LARGE Compact fields and mass generation}
\footnote{The work has been supported by the RFBR grant 99-01-01230.}

\vspace*{0.5cm}
{\large
V.K.~Mitrjushkin
}\\

\vspace*{0.2cm}
{\normalsize
{\em Lab. Theor. Phys.,
Joint Institute for Nuclear Research, 141980 Dubna, Russia}
}

\vspace*{0.5cm}
{\bf Abstract}
\end{center}

It is shown that the free propagator of an angular, i.e. compact, 
field with zero lagrangian mass acquires a nonzero propagator mass 
$~\omega~$ (`kinematical' mass generation).
To demonstrate this effect the free propagator of the 
goldstone boson in an $~O(2)~$ model with spontaneous symmetry
breaking is calculated. It is shown that this propagator is massive,
the mass $~\omega~$ being a function of the scalar `condensate' 
$~{\bar\phi}~$.

\section{Introduction}\setcounter{equation}{0}

Compact functional integrals appear in a number of field theoretical and
statistical models. For instance, all integrals in lattice gauge
theories \cite{wils} are compact as well as integrals in nonlinear 
sigma-models, etc..  One interesting example is connected with the 
propagator of the goldstone boson in a theory with spontaneously 
symmetry breaking.

Let us consider a simple $d$--dimensional field theoretical model with
$~O(2)\sim U(1)~$ global symmetry \cite{gold}  (see
also \cite{gosw},\cite{guhk}) Throughout this paper a lattice
regularization will be used and $~d\ge 3$.  Let $~\phi_x = \phi^{(1)}_x + 
i\phi^{(2)}_x~$ be a complex scalar field defined on the infinite 
lattice where components $~\phi^{(i)}_x~$ ($i=1,2)$ are noncompact, 
i.e. $~-\infty \le\phi^{(i)}_x \le\infty~$.  The action $~S~$ is given 
by

\eq
S = \sum_x \left[ \frac{1}{2}\sum_{\mu} \partial_{\mu}\phi_x^{\ast}
\cdot \partial_{\mu}\phi_x - \frac{m^2}{2}\phi_x^{\ast}\phi_x
+ \lambda (\phi_x^{\ast}\phi_x)^2 \right]~,
                 \label{action}
\en

\noi where $~m^2 > 0~$,
$~\partial_{\mu}\phi_x = \phi_{x+\mu}-\phi_x~$ and
the lattice spacing is chosen to be unity.
Evidently, action $~S~$ is invariant with respect to the
transformations $~\phi_x\to e^{i\alpha}\phi_x~$ where $~\alpha~$
is some constant. This action has an infinite number of minima 
(`vacua') at $~\phi_x ={\bar \phi}\cdot e^{i\theta}~$ where
$~{\bar \phi}^{\, 2} = m^2/4\lambda~$ and $~-\pi < \theta \le \pi~$.
The average of any functional $~\cao(\phi)~$ is

\eq
\Bigl\langle \cao(\phi)\Bigr\rangle = \frac{1}{Z}
\int_{-\infty}^{\infty}\!\prod_zd\phi^{(1)}_z\phi^{(2)}_z 
~\cao(\phi) \cdot e^{-S(\phi)} ~,
              \label{average_O}
\en

\noi and the partition function $~Z~$ is defined by
$~\langle 1\rangle =1~$. 
To trace the appearence of the goldstone boson it is convinient to make 
a change of variables
$~\phi_x = \rho_x \cdot e^{\theta_x}~$ where $~0\le\rho_x \le \infty~$
and $~-\pi < \theta_x\le \pi~$.
It is important to note that ranges of the variables 
 $~\rho_x~$ and $~\theta_x~$ do not extend over the whole real axis.
Evidently, the main contribution to the integral in eq.(\ref{average_O})
comes from the region $~\rho_x \simeq {\bar\phi}~$,
at least, for large enough $~m^4/\lambda~$.
Making the change of variables
$~\rho_x =\rho_x^{\prime} + {\bar\phi}~$ one observes that
new radial field $~\rho_x^{\prime}~$ has a mass term 
$~m_{\rho^{\prime}}^2\sum_x\rho_x^{\prime\, 2}~$ 
where $~m_{\rho^{\prime}}^2 = 8\lambda {\bar\phi}^2~$,
and angle field $~\theta_x~$ has no mass term (goldstone boson). 
Rescaling $~\theta$--field ($\theta_x \equiv
\varphi_x/{\bar\phi}$) one finds the free action of the (massless) 
$\varphi$-field~:

\eq
S_0(\varphi) = \frac{1}{2}\sum_{x\mu} (\partial_{\mu}\varphi_x)^2~;
\qquad
|\varphi_x| \le M \equiv \pi{\bar\phi}~.
\en

\noi This field is compact and the corresponding free propagator 
$~\tG_{xy}~$ is given by

\eq
\tG_{xy}  = \Bigl\langle \varphi_x\varphi_y \Bigr\rangle_0
\equiv \frac{1}{Z_0}\int_{-M}^{M}\!\prod_zd\varphi_z 
~\varphi_x\varphi_y \cdot e^{-S_0(\varphi)}~,
            \label{free_prop}
\en

\noi where $~Z_0~$ is defined by $~\langle 1\rangle_0 =1~$.
The problem is to calculate this integral at finite $~M~$.
The same problem arises in perturbative expansion in lattice gauge 
theories, nonlinear sigma--models, XY--model and others.

It is the aim of this paper to calculate the free propagator
defined in eq.(\ref{free_prop}) at $~M < \infty~$.
The next section is dedicated to the integration identities method 
which permits to calculate compact functional integrals. 
In the third section the analytical expression for the free correlator 
$~\tG_{xy}~$ at large $~M~$ is given.  Fourth section is dedicated to 
the numerical calculation of the propagator $~\tG(p)~$ at nonzero 
momenta and the comparison with analytical results.  The last section 
is reserved for summary and discussions.

\section{Integration identities}\setcounter{equation}{0}

Let  $~\cao(\varphi)~$ be any functional of the compact field
$~\varphi_x~$ and

\eq
\Bigl\langle \cao \Bigr\rangle_0 = \frac{1}{Z_0} \int_{-M}^{M}
\!\prod_zd\varphi_z ~\cao(\varphi) \cdot e^{-S_0(\varphi)}~.
            \label{avo}
\en

\noi Let us make an infinitesimal change of variables

\eq
\varphi_x \to \varphi^{\prime}_x ~=~ \varphi_x-\delta f_x~,
            \label{change_of_var_1}
\en

\noi where $~\delta f_x~$  could be {\it any} infinitesimal
parameters and

\eq
-M-\delta f_x \le \varphi^{\prime}_x \le M-\delta f_x~.
\en

\noi Then

\eq
\cao(\varphi) = \cao(\varphi^{\prime}) + \delta\cao(\varphi^{\prime})
\equiv \cao(\varphi^{\prime}) +
\sum_x \delta f_x \cdot \delta\cao_x(\varphi^{\prime}) +\ldots~,
\en

\noi and

\eqa
S_0(\varphi) &=& \frac{1}{2}\sum_{x\mu} (\partial_{\mu}\varphi_x)^2
~\equiv~  S_0(\varphi^{\prime}) + \delta S_0(\varphi^{\prime})~;
\nonumber \\
\nonumber \\
\delta S_0(\varphi^{\prime}) &=& \sum_{x} \delta f_x
\cdot \Delta\varphi^{\prime}_x + \ldots~,
\ena

\noi where $~\Delta = -\sum_{\mu}\partial_{\mu}\bpartial_{\mu}~$
and $~\bpartial_{\mu}\varphi_x = \varphi_x - \varphi_{x-\mu}~$.

In the case of finite $~M~$ the variation of the limits of 
integration must be taken into account. For any functional
$~\Phi(\varphi)~$ one obtains

\eq
\int_{-M}^{M}\! d\varphi_x ~\Phi(\varphi)
~=~ \int_{-M-\delta f_x}^{M-\delta f_x}\! 
d\varphi_x^{\prime} ~\Phi^{\prime}(\varphi^{\prime}) 
= \int_{-M}^{M}\! d\varphi_x^{\prime} ~\Phi^{\prime}
+ \delta^{\prime}\int_{-M}^{M}\! d\varphi_x ~\Phi^{\prime}(\varphi)~,
\en

\noi where

\eq
\delta^{\prime}\int_{-M}^{M}\! d\varphi_x ~\Phi^{\prime}(\varphi)
= -\delta f_x \int_{-M}^{M}\! d\varphi_x^{\prime} 
~\Phi^{\prime} \cdot \Bigl[ \delta (\varphi_x^{\prime}-M)
- \delta (\varphi_x^{\prime}=-M)\Bigr] +\ldots ~.
\en

\noi Therefore,

\eq
\delta^{\prime}\Bigl\langle \cao\Bigr\rangle_0 \equiv
-\sum_x\delta f_x\cdot\left\langle \Bigl[ \delta(\varphi_x-M)
-\delta(\varphi_x+M)\Bigr] \cao \right\rangle_0 + \ldots ~.
\en

\noi Finally, one obtains the integration identities

\eq
\delta\Bigl\langle \cao\Bigr\rangle_0
\equiv \Bigl\langle \delta\cao\Bigr\rangle_0
- \Bigl\langle \cao\cdot\delta S \Bigr\rangle_0
+ \delta^{\prime}\Bigl\langle \cao\Bigr\rangle_0 ~\Ra~ 0~
\en

\noi or

\eq
\Bigl\langle \delta\cao\Bigr\rangle_0 - \sum_x \delta f_x 
\left\{ \Delta_x \Bigl\langle \varphi_x\cao \Bigr\rangle_0
+ \left\langle \Bigl[ \delta(\varphi_x-M)
-\delta(\varphi_x+M)\Bigr] \cao \right\rangle_0 \right\} ~=~ 0~.
             \label{wt_1}
\en

\noi In the case of noncompact integrals, i.e. $M=\infty$,
one obtains standard (Ward--Takahashi) identities

\eq
\Bigl\langle \delta\cao\Bigr\rangle_0 - \sum_x \delta f_x 
\cdot \Delta_x \Bigl\langle \varphi_x\cao \Bigr\rangle_0 ~=~ 0~.
             \label{wt_2}
\en

\section{Free propagator}\setcounter{equation}{0}

Let us choose $~\cao = \varphi_y~$. Then
$~\delta\cao = \delta f_y =\sum_x \delta f_x \cdot \delta_{xy}~$
and from eq.(\ref{wt_1}) one obtains

\eq
\Delta_x \tG_{xy} + \left\langle \Bigl[ \delta(\varphi_x-M)
-\delta(\varphi_x+M)\Bigr] \varphi_y \right\rangle_0 ~=~\delta_{xy}~,
                    \label{wt_3} 
\en

\noi where

\eq
\tG_{xy} \equiv \Bigl\langle \varphi_x \varphi_y \Bigr\rangle_0
~=~ \frac{1}{(2\pi)^d} \int_{-\pi}^{\pi}\!d^dp~
e^{ip(x-y)} \cdot \tG(p)~.
\en

\noi If $~M=\infty~$ then from eq.(\ref{wt_2}) follows

\eq
\Delta_x G_{xy} = \delta_{xy}~,
\en

\noi and the solution is

\eq
G_{xy} = \frac{1}{(2\pi)^d} \int_{-\pi}^{\pi}\!d^dp~
\frac{e^{ip(x-y)} }{\cak^2(p)}~;
\qquad
\cak^2(p) =\sum_{\mu}4\sin^2 \frac{p_{\mu}}{2}~.
                 \label{prop_stan}
\en

\noi This propagator is well--defined at $~d\ge 3~$. Let $~J_x~$ be 
some current. Then

\eqa
\Bigl\langle e^{iJ\varphi} \Bigr\rangle_0 &=&
\exp \left\{-\frac{1}{2} J\tG J + \delta F(J;M) \right\}~;
          \label{curr}
\\
\nonumber \\
\delta F(J;M) &=& \sum_{n\ge 2} \delta F^{(n)}(J;M)~;
\nonumber \\
\delta F^{(n)}(J;M) &=& \sum_{\{ x^{(i)} \} }
\delta F^{(n)}_{x^{(1)}\ldots x^{(2n)}}(M)
\prod_{i=1}^{2n}J_{x^{(i)} }~.
\nonumber
\ena

\noi At $~d\ge 3~$

\eq
\tG_{xy}\to G_{xy}~; \qquad
F^{(n)}_{x^{(1)}\ldots x^{(2n)}}(M) \to 0~
\en

\noi  when $~M\to \infty~$. Therefore, $~\delta F(J;M)~$ is a small
correction at large $~M~$.

\noi Let us calculate the matrix element in eq.(\ref{wt_1}).

\eq
\left\langle \Bigl[ \delta(\varphi_x-M)
-\delta(\varphi_x+M)\Bigr] \varphi_y \right\rangle_0
= -\frac{1}{\pi}\int_{-\infty}^{\infty}\!d\xi ~
\sin (\xi M) \cdot \frac{\partial}{\partial J_y}
\left\langle e^{iJ\varphi} \right\rangle_0\Big|_{J_z=\xi\delta_{zx}}~.
\en

\noi At large enough $~M~$ one can discard $~\delta F~$ in eq.(\ref{curr})
and 

\eq
\left\langle \Bigl[ \delta(\varphi_x-M)
-\delta(\varphi_x+M)\Bigr] \varphi_y \right\rangle_0
= \omega^2 \cdot \tG_{xy}~,
\en

\noi where

\eq
\omega^2 = \frac{M\tzeta^3\sqrt{2}}{\sqrt{\pi}}\cdot 
e^{-\frac{1}{2}\tzeta^2M^2}~,
                     \label{omega2}
\en

\noi and

\eq
\tzeta^2 = \frac{1}{\tG_0}~;
\qquad \tG_0 \equiv \tG_{xx}~.
               \label{cond_zeta}
\en

\noi Finally, one obtains

\eq
\Bigl( \Delta_x + \omega^2 \Bigr) \tG_{xy} = \delta_{xy}~,
\en

\noi and the propagator $~\tG~$ is given by

\eq
\tG_{xy} = \frac{1}{(2\pi)^d} \int_{-\pi}^{\pi}\!d^dp
\frac{e^{ip(x-y)} }{\cak^2(p) + \omega^2}~,
                 \label{prop}
\en

\noi  where $~\tzeta~$ should satisfy

\eq
\frac{1}{\tzeta^2} = \frac{1}{(2\pi)^d} \int_{-\pi}^{\pi}\!d^dp~
\frac{1}{\cak^2(p) + \omega^2}~.
\en

\noi It is easy to see that in the limit  $~M\to\infty~$ 
$~\tzeta\to\zeta~$ where $~\zeta~$ is given by

\eq
\frac{1}{\zeta^2} \equiv \frac{1}{(2\pi)^d} \int_{-\pi}^{\pi}\!d^dp~
\frac{1}{\cak^2(p)}~.
\en

\section{Numerical calculation of $~\tG(p)~$ }\setcounter{equation}{0}

Results of the analytical calculation of the propagator $~\tG~$ 
defined in eq.(\ref{free_prop}) can be 
confronted with the results of the numerical study of this propagator 
on a final lattice.
In the previous section this propagator has been analytically studied 
at large values of $~M~$, so that the mass squared $~\omega^2(M) \ll 
1~$. On the other side, method Monte Carlo gives a possibility to study 
this propagator also at small values of $~M~$.

The Fourier transform of the propagator 

\eq
 \Bigl\langle \varphi(p)\varphi(-p) \Bigr\rangle_0
= \frac{1}{Z_0}\int_{-M}^{M}\!\prod_zd\varphi_z 
~\varphi(p)\varphi(-p) \cdot 
e^{-\sum_{x\mu} (\partial_{\mu}\varphi_x)^2}
= \frac{V}{2}\tG(p)  
            \label{free_prop_mom}
\en

\noi was calculated numerically on the finite lattice at different
values of $~M~$ and momenta $~p~$, $~V~$ being a number of sites.
It is convinient to define the effective mass $~\omega_{\eff}(p)~$ :

\eq
\omega^2_{\eff}(p) = \tG^{-1}(p) - \cak^2(p)~.
\en

\noi Evidently, the propagator $~\tG(p)~$ is massive if 
$~\omega^2_{\eff}(p)~$ does not depend on $~p~$.

\begin{figure}[pt]
%\vspace{1.0cm}
\begin{center}
%\vskip -1.5truecm
\leavevmode
\hbox{
\epsfysize=14cm
\epsfxsize=14cm
\epsfbox{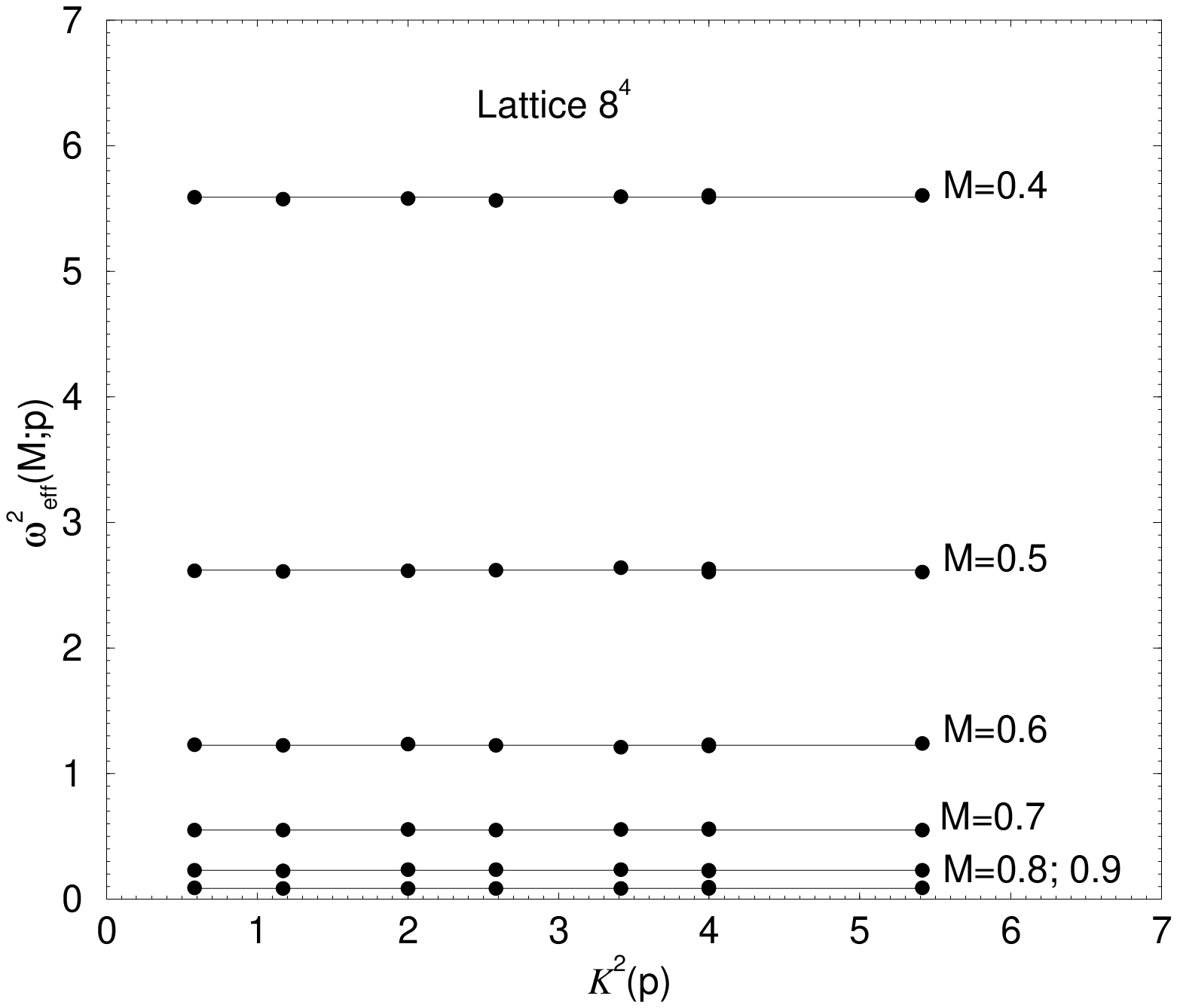}
}
\end{center}
%\vskip 1.5truecm
\caption{Dependence of $\omega^2_{\eff}(M;p)$ on $\cak^2(p)$ for different
$M$. Solid lines correspond to averages $\omega^2(M)$.
}
\label{fig:om1_4d_08x08}
%\vskip -0.5truecm
\end{figure}

In Figure \ref{fig:om1_4d_08x08} the dependence of the effective mass 
$~\omega^2_{\eff}(p)~$ on $~\cak^2(p)~$ for different $~M~$ is shown.  
Solid lines correspond to averages of $~\omega^2_{\eff}(p)~$ for every 
$~M~$. One can see that the dependence on $~p~$ is practically absent
as it was expected.

\begin{figure}[pt]
%\vspace{1.0cm}
\begin{center}
\vskip -1.5truecm
\leavevmode
\hbox{
\epsfysize=14cm
\epsfxsize=14cm
\epsfbox{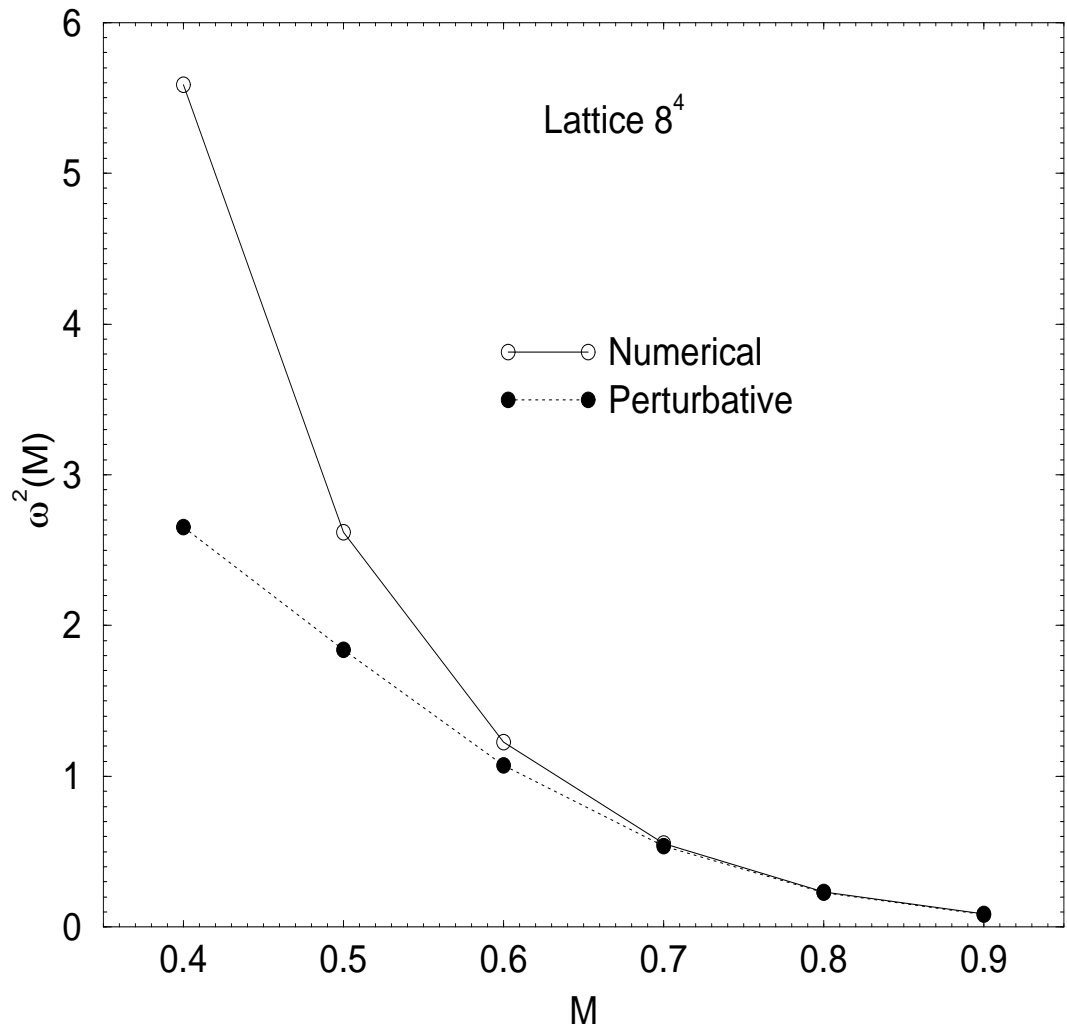}
}
\end{center}
%\vskip 1.5truecm
\caption{Dependence of $\omega^2(M)$ on $M$.
Lines are to guide the eye.
}
\label{fig:om2_4d_08x08}
%\vskip -0.5truecm
\end{figure}

It is interesting to compare numerically calculated 
$~\omega^2=\omega^2(M)~$ with that given in eq.(\ref{omega2}) (taking
into account $~M\to M{\sqrt 2}~$ because of the different normalization
of the action in eq.(\ref{free_prop_mom})~). 
In Figure \ref{fig:om2_4d_08x08} one can see that at $~M
\ageq 1~$ the results of analytical calculation (full circles) converge
to $~\omega^2~$ calculated numerically (open circles).

\section{Summary and discussions}\setcounter{equation}{0}

It is shown that the free propagator of an angular, i.e. compact, field with 
zero lagrangian mass acquires at $~d\ge 3~$ a nonzero 
dimensionless propagator mass $~\omega~$ (kinematical mass 
generation).

To demonstrate this effect the free propagator of the goldstone boson 
in an $~O(2)~$ lattice model with spontaneous symmetry breaking is 
calculated. It is shown that this propagator is massive, the mass 
$~\omega~$ being a function of the scalar `condensate' $~{\bar\phi}~$.  
The results of numerical simulations on the finite lattice are in good 
agreement with the results of analytical calculations.

In the case of spin systems lattice has a physical meaning and no 
continuum limit is needed. In the case of continuum theory
\footnote{It is worthwhile to note that $~\lambda\phi^4~$ theory in 
four dimensions has no nontrivial continuum limit \cite{triv}.}
the question of interest is the dependence of the dimensional mass
$~m_{dim}~$ $$ m_{dim} = \frac{\omega(M)}{a}~$$ in the limit $~a\to 0~$
where $~a~$ is lattice spacing.

 The appearence of a nonzero (propagator) mass $~\omega~$, i.e.  the 
effect of kinematical mass generation, could help to resolve the 
problem of the strong infrared divergencies in $~4d~$ QCD at finite
temperature \cite{lin1}. Indeed, in QCD with lattice regularization 
all integrals are compact. Therefore, $~m_{dim}(a) = \omega/a ~$ is 
nonzero and provides a natural regularization of the infrared 
divergencies at finite spacing $~a~$. Finally, it can give rise to 
the infrared cutoff discussed in \cite{lin1}.

In principle, one can speculate that the mechanism of the kinematical
mass generation gives a possibility to obtain massive gauge fields
without the introduction of higgs bosons.

\newpage
%%%%%%%%%%%%%%%%%%%%%%%%%%%%%%%%%%%%%%%%%%%%%%%%%%%%%%%%%%%%%%
%%%%%%%%%%%%%%%%%%%%%%  Figure 1  %%%%%%%%%%%%%%%%%%%%%%%%%%%%
%%%%%%%%%%%%%%%%%%%%%%%%%%%%%%%%%%%%%%%%%%%%%%%%%%%%%%%%%%%%%%

%
%Now comes Figure : "om1_4d_08x08.xm"
%

\vfill

%%%%%%%%%%%%%%%%%%%%%%%%%%%%%%%%%%%%%%%%%%%%%%%%%%%%%%%%%%%%%%
%%%%%%%%%%%%%%%%%%%%%%  Figure 2  %%%%%%%%%%%%%%%%%%%%%%%%%%%%
%%%%%%%%%%%%%%%%%%%%%%%%%%%%%%%%%%%%%%%%%%%%%%%%%%%%%%%%%%%%%%

%
%Now comes Figure : "om2_4d_08x08.xm"
%

\vfill

\end{document}